\documentclass[11pt,a4paper]{article}
\usepackage[all]{xy}
\usepackage{amsmath,amsfonts,amssymb,amsthm,amscd,pb-diagram,pb-xy}
\usepackage{graphicx}
\usepackage{epsfig}
\usepackage{caption}

\usepackage{color}  

\definecolor{red-}{rgb}{0.8,0.0,0.0}

\definecolor{green-}{rgb}{0.0,0.7,0.0}

\definecolor{blu-}{rgb}{0.0,0.0,1.0}

\newenvironment{Remark}{\par \medskip \noindent{\sc Remark:}}

\newcommand{\Reali}{\ensuremath{\mathbb{R}}}

\def\vth{\vartheta}

\def\A{\mathcal A\/}
\def\B{\mathcal B\/}

\def\M{\mathcal M\/}

\def\S{\mathcal S\/}

\def\J{J_1\/(\M)}

\def\JS{J_1\/(\S)}

\def\={\, = \,}
\def\de#1/de#2{\frac{\partial {#1}}{\partial {#2}}}
\def\De#1/de#2{\dfrac{\partial {#1}}{\partial {#2}}}

\def\pl{{\mathbf P}_L}
\def\pr{{\mathbf P}_R}
\def\p{{\mathbf P}}

\begin{document}

	\title{An alternative approach to the Painlev\'e paradox through constitutive characterization of constraints in impulsive mechanics}

	\author{Stefano Pasquero\thanks{The author thanks the support by University of Parma and by the Italian National Group of Mathematical Physics
			(GNFM-INdAM)}\\
		Departments of Mathematics, Physics and Computer Sciences \\
		University of Parma \\
		Parco Area delle Scienze 53/a, 43124 PARMA -- Italy \\
		E-mail: stefano.pasquero@unipr.it
	}
	\date{}
	\maketitle
	\begin{abstract}
		\noindent
		We frame the Painlevé mechanical system, which has been extensively studied because of the paradox it generates, within the class of Regular Geometric Impulsive Mechanical Systems (RGIMS), by modeling it as a mechanical system subject to a rough unilateral positional constraint
		$\S$, where friction is represented by an instantaneous kinetic constraint
		$\B$, internal to $\S$ and of impulsive nature.

		\noindent The evolution of the system is therefore determined by the choice of a constitutive characterization for these constraints, a choice that restores mechanical determinism and eliminates any paradoxical aspects of the system’s behavior, in agreement with experimental evidence.

		\noindent It is shown that, similarly to what occurs in general non--ideal impulsive systems, the choice of a constitutive characterization of the constraint system depends on the determination of two numerical coefficients $\sigma$ and $\beta$, which depend on the kinematic and mass-related data of the system, and possibly also on physical quantities not strictly of a mechanical nature, such as material properties. The simplicity of the model also allows for a straightforward experimental analysis of the system’s behavior and for the experimental determination of the values of these coefficients.
	\end{abstract}

	\section*{Introduction}

The formal study of Classical Mechanics, as is well known, is based on a number of axiomatic principles that enable the construction of a theoretical model whose predictions must be subjected to experimental verification. Among these, the following are of fundamental importance:
\begin{itemize}
	\item \textbf{Newton’s Second Law of Dynamics}, heuristically summarized by the relation
	$\mathbf{F} = m\mathbf{a}$
	which applies to each individual material point of the system and establishes a relationship between the forces acting on the point and its acceleration;

	\item \textbf{The Principle of Causality}, heuristically expressed by stating that, within the law of dynamics, forces are the \emph{cause} of accelerations, and accelerations are the \emph{effect} of forces;

	\item \textbf{The Principle of Determinism}, heuristically expressed by stating that the mechanical evolution of a system can always be uniquely determined from the knowledge of the forces acting on it and of the initial data of position and velocity of the system itself.
\end{itemize}

A direct consequence of the principle of causality, when analyzing a purely mechanical problem, consists in determining the structure of the dynamical law in the form $	\mathbf{F}(t, P, \mathbf{v}) = m\mathbf{a}$
where the forces constitute given data of the problem and may depend on time, position, and velocity of the point, but not on the acceleration itself.

Moreover, the three aforementioned principles imply that it is not possible for the mechanical evolution of the point, once the acting forces and the initial position and velocity are known, to either fail to exist or to exist in two or more distinct forms.

These three principles and their consequences find, on the one hand, experimental confirmation in a wide range of examples (as long as one remains within the domain of applicability of Classical Mechanics, understood as a local physical theory), and on the other hand, mathematical confirmation in Cauchy’s theorem on existence and uniqueness of the solution of a system of second-order differential equations written in normal form.

\smallskip

The same principles, or analogous ones suitably adapted to the context, can be introduced in the study of impulsive Classical Mechanics. In this case, the second law of dynamics takes the integral form $
\mathbf{I}(t, P, \mathbf{v}_L) = m \mathbf{v}_R$,
where $\mathbf{I}$ is the instantaneous impulse acting on the point---assumed to be a known datum of the problem once the initial conditions are assigned---and $\mathbf{v}_L, \mathbf{v}_R$ are the velocities of the point before and after the action of the impulse, respectively. This formulation, unlike the usual expression $\mathbf{I} = m \Delta \mathbf{v}$, explicitly accounts for the principle of causality.

In impulsive mechanics, the principle of determinism remains essentially unchanged with the concept of impulse replacing that of force.

\smallskip

Differently from what occurs in the study of the motion of one or more ``free'' material points, the study of mechanical systems subject to constraints, that are prescriptions concerning admissible positions and/or velocities of the system, reveals a violation of the determinism principle. Indeed, by their very nature, constraints must interact with the mechanical system through forces or impulses, called \emph{reactive} to distinguish them from the \emph{active} ones considered so far, in such a way as to compel the system to evolve only within the set of admissible configurations and velocities. This must hold regardless of the system of active forces or impulses acting upon it.

Consequently, the dynamic law assumes the form $\mathbf{F} + \boldsymbol{\Phi} = m\mathbf{a}$ with the reactive forces $\boldsymbol{\Phi}$  unknown quantities of the mechanical problem, in addition to those determining the motion of the system (the same holds in the case of an impulsive system). In fact, while the constraint prescriptions reduce the number of parameters needed to determine the motion (the so-called degrees of freedom of the system), the reactive forces and impulses introduce an equal number of unknowns corresponding to the scalar (differential or integral) equations of Dynamics.

\smallskip

To restore the validity of the determinism principle, it is therefore necessary to impose a \emph{constitutive characterization of the constraint}, namely a prescription---valid for any kinematic state and any system of active forces or impulses---on which reactive forces or impulses the constraint may or may not exert on the mechanical system, under the sole condition of reinstating determinism. Of course, in addition to being formally admissible in the sense just stated, for such a constitutive characterization to be physically meaningful, it must withstand experimental verification.

\smallskip

In light of what has been recalled so far, the mechanical system determining the so--called \emph{Classical Painlev\'e Paradox} appears, in all respects, to generate a genuine paradox. Recall that the paradox appears when studying the motion of a homogeneous rigid rod, unilaterally constrained to move along the upper side of a rough plane satisfying the usual laws of dry friction (see Fig. 1). Given suitable initial conditions for the motion of the rod, with one of its ends in contact with the plane, and for specific values of the friction coefficient, the resulting equations of motion become non-deterministic: multiple admissible solutions may exist, or no solution at all.

\medskip

Two remarks in this regard are in order. The first concerns the experimental aspects of the paradox. It is unreasonable to assume that, in a laboratory experiment devised to reproduce Painlev\'e mechanical problem, the time evolution of the rod would fail to exist, or that the rod would evolve randomly following one or another of the possible solutions. Indeed, in various possible experimental realizations of the problem—some even quite elementary—the rod always evolves in the same way (compatible with experimental precision).

The second remark concerns the logical aspects of the paradox. When a fundamental principle of a theory appears to be violated, the alternatives are either to abandon the theory founded upon that principle, or to verify whether additional hypotheses have been introduced that are incompatible with the adopted principles. The second possibility is clearly the case in the Painlev\'e paradox: in addition to the usual principles, an arbitrary constitutive characterization of the unilateral rough constraint has been chosen, namely that provided by the dry friction laws. Before labeling the problem as paradoxical, it is therefore reasonable to abandon that particular constitutive assumption and to search instead for one that, first, does not lead to violations of foundational principles and, second, that exhibits good agreement between theoretical predictions and experimental evidence.

\medskip

Before addressing the more technical aspects of this work, it is necessary to recall one more theoretical aspect: the initial data of a mechanical problem can heavily influence the nature of the problem itself.

Consider the elementary example of a material point moving in the upper half-plane, bounded by a horizontal straight line. Suppose the initial data place the point in contact with the line, with initial velocity $\mathbf{v}_0$. The relationship between $\mathbf{v}_0$ and the normal vector $\mathbf{N}$ to the line (oriented upwards) determines the type of analysis required for the problem. Specifically:
\begin{itemize}
	\item[i)] if $\mathbf{v}_0 \cdot \mathbf{N} > 0$, the velocity $\mathbf{v}_0$ represents the initial velocity with which the point \emph{departs} from the boundary line. The point immediately detaches from the line, and the system possesses two degrees of freedom, with its evolution governed by Newton’s second law
	$\mathbf{F}(t, P, \mathbf{v}) = m \mathbf{a}$;

	\item[ii)] if $\mathbf{v}_0 \cdot \mathbf{N} < 0$, the velocity $\mathbf{v}_0$ represents the velocity with which the point \emph{reaches} the boundary line. It must therefore be regarded as the $\mathbf{v}_L$ of an impulsive system, and the evolution of the point is determined by the integrated form of Newton’s law,
	$\mathbf{I}(t, P, \mathbf{v}_L) = m \mathbf{v}_R$,
	where the constitutive characterization of the line as a \emph{unilateral impulsive constraint} becomes relevant;

	\item[iii)] if $\mathbf{v}_0 \cdot \mathbf{N} = 0$, the velocity $\mathbf{v}_0$ may be regarded either as the speed with which the point departs from the initial position or as the speed with which it reaches it. It is well known that, in the case of a smooth constraint, the two velocities coincide, the system is non-impulsive, and its evolution---at least in a neighborhood of the initial instant---is again governed by Newton’s second law. In the case of a rough constraint, however, the choice may depend on the constitutive characterization of the line.
\end{itemize}

In this article, assuming the three principles listed above to be inviolable, we propose an approach to the study of the Painlevé system that relies exclusively on the law of Impulsive Mechanics  for constrained systems, formulated in terms of instantaneous jumps of velocities and impulse-based interactions,
rather than on the law of Classical Mechanics, formulated in terms of time evolution and force-based interactions.

We frame the Painlev\'e mechanical system within the differential-geometric context of Regular Geometric Impulsive Mechanical Systems (RGIMS). Within this framework, the roughness of the plane is modeled by introducing an ``internal'' kinetic constraint, so that the system is subject to two constraints: the positional constraint associated with the rough plane, and the kinetic constraint associated with surface friction. This modeling naturally leads to two geometrically privileged directions: one orthogonal to the rough plane, and another parallel to the plane but orthogonal to the kinetic constraint induced by friction.

The analysis of the initial data corresponding to the manifestation of the paradox shows that, at the initial instant, the system possesses zero velocity in the direction orthogonal to the rough plane, but a nonzero velocity component in the direction orthogonal to the kinetic constraint. Consequently, the evolution of the system near the initial data is governed by the laws of impulsive Mechanics rather than by the ones of ``smooth''  Mechanics. Moreover, the assignment of an impulsive characterization of the constraints is mandatory to restore determinism.

In the absence of additional directions dictated by supplementary principles (possibly dependent on the forces acting on the system), the simplest possible assignment of a reactive impulse acting on the system is a linear combination of the two possible directions, with coefficients depending on the massive, geometric and kinetic data of the system. Such a constitutive characterization, that is the simplest possible
within the proposed model, has several advantages: it avoids any paradox associated with the principle of determinism; it exhibits, at least in principle, qualitative agreement with experimental data; it allows an experimental evaluation of the coefficients of the linear combination.

\smallskip

This article is divided into three sections. The first is devoted to a brief presentation of the classical approach to the study of the dynamics of the Painlevé mechanical system and to the origin of the paradox. It should be regarded solely as an introduction to the general problem and it is not, and cannot be, exhaustive on the subject, which is still the focus of ongoing analysis and research. Readers interested in the traditional formulation of the Painlev\'e paradox are referred to the extensive literature on the subject, such as \cite{Brogliato, GILARDI2002, pfeiffer1995impacts, Stewart2000,GenotBrogliato1999,champneys2016painleve} and the references therein.

The second section is divided into two subsections. In the first, the Painlevé system is framed within the geometric setting of RGIMS, and the directions orthogonal to the unilateral positional constraint and to the instantaneous kinematic constraint are exhibited. In the second, the constitutive characterization of the constraint system is described from the standpoint of general theory, and several possible examples of constitutive characterization are presented.

The third section briefly summarizes the conclusions of the paper and outlines some possible developments of the proposed method.

\smallskip

Given the specific nature of the adopted approach, the bibliography is intentionally minimal, providing only the essential references needed for self-consistency.

	\section{The classical approach to the Painlev\'e system}

The usual description of the Painlev\'e paradox begins with a mechanical system consisting of a homogeneous rod of length $2L$, mass $m$, and moment of inertia $A$, moving in the upper vertical half-plane bounded by a rough horizontal line.
\begin{center} \label{FigRod}
	\vskip -0.1truecm \begin{minipage}[l]{.50\textwidth}
		The system therefore has three degrees of freedom, denoted by $(x, y)$ --- the coordinates of the center of mass $G$ of the rod --- and by $\vartheta$, the orientation of the rod with respect to the horizontal axis.

		The rod is subjected to the active force of gravity $m {\mathbf g}$ and, since at the initial time $t_0$ the rod has its endpoint $P$ in contact with the horizontal line, it is also subjected to the reactive force ${\mathbf \Phi}$ (see Fig.~1).
	\end{minipage}%
	\begin{minipage}[r]{.50\textwidth}
		\begin{center}
			\hskip0.8truecm	\includegraphics[width=0.8\textwidth]{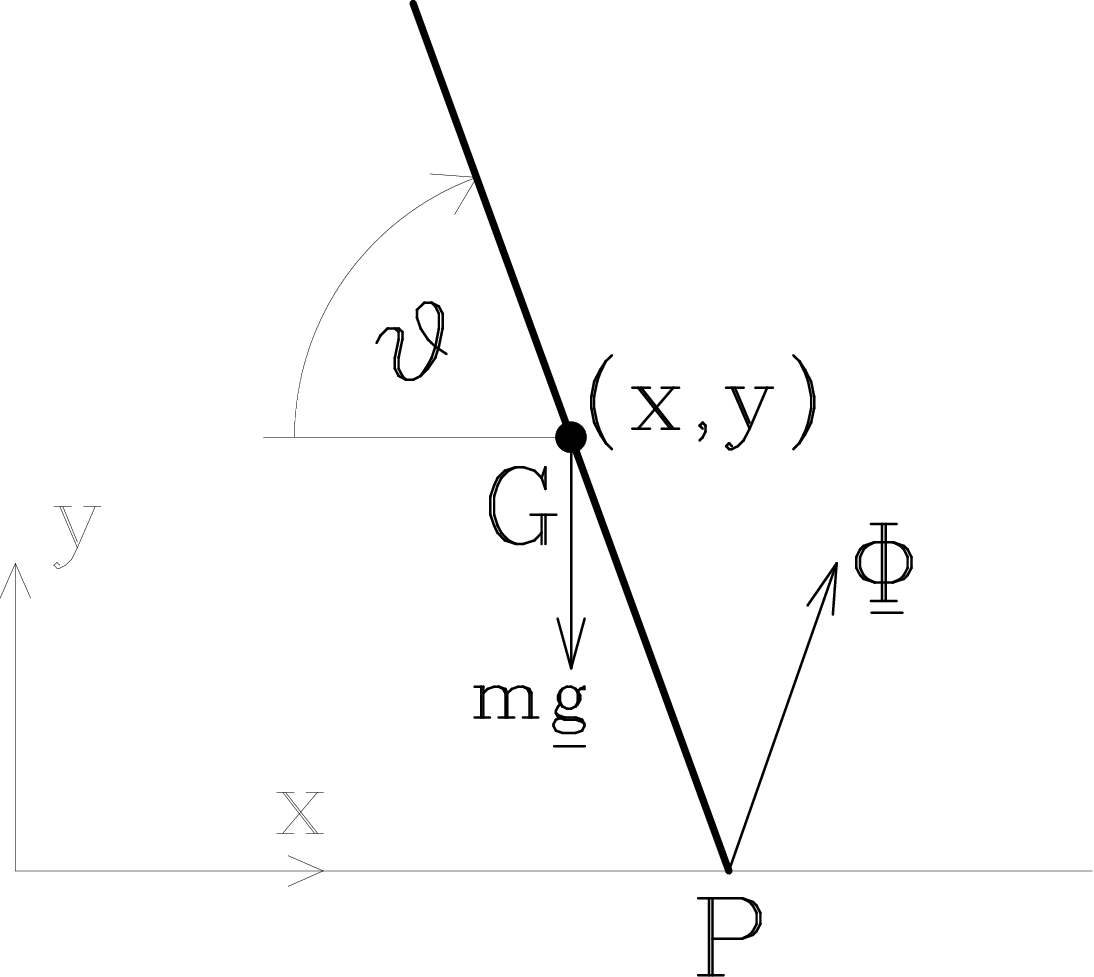}
			\textrm{Fig. 1}
		\end{center}
	\end{minipage}
\end{center}

Let $\mathbf{e}_x, \mathbf{e}_y$ the unit vectors parallel and orthogonal, respectively, to the rough line, and ${\mathbf u} \= \frac{1}{L} \, \vec{PG}$ the unit vector having the direction of the rod.
With this choice of coordinates and axes, the velocity of the contact point $P$ is
\begin{eqnarray}\label{velPuntoContatto}
	\mathbf{v}_P = \left(\dot{x} - L \dot{\vartheta} \sin \vartheta \right)\mathbf{e}_x
	+ \left(\dot{y} - L \dot{\vartheta} \cos \vartheta \right)\mathbf{e}_y \, .
\end{eqnarray}
For later use, and for brevity's sake, we will say that the rod is pushed if $(\mathbf{v}_P \cdot \mathbf{e}_x) \cos\vth >0$, and it is pulled if $(\mathbf{v}_P \cdot \mathbf{e}_x) \cos\vth < 0$.

The motion at the initial time $t_0$ is such that the velocity $\mathbf{v}_P(t_0)$ of $P$ is tangent to the line itself, so that, with obvious notation, $\mathbf{v}_P(t_0) = (\dot{x}_0 - L \dot{\vartheta}_0 \sin \vartheta_0)\mathbf{e}_x$.

\smallskip

The reaction of the rough line ${\mathbf \Phi} = \Phi_x \mathbf{e}_x + \Phi_y \mathbf{e}_y$ satisfies the Amontons–Coulomb constitutive characterization of dry friction:
\begin{subequations}\label{AC}
	\begin{eqnarray}\label{ACS}
		\hskip -0.4truecm | { \Phi}^s_x | \, &\le \, \mu_s |
		{ \Phi}^s_y | \quad\quad\quad\quad\quad\quad\quad\quad\,\,\, {\rm if} \,\, {\bf v}_P \= 0
	\end{eqnarray}   \vskip-0.7truecm
	\begin{eqnarray}\label{ACD}
		{\Phi}^d_x \, \, = \, -  \mu_d  \, |{\Phi}^d_y | \,
		sign\left(\dfrac{{\bf v}_P}{\| {\bf v}_P \|} \right) \qquad {\rm if} \,\, {\bf v}_P \ne 0
	\end{eqnarray}
\end{subequations}
where the superscripts $s$ and $d$ refer to static and dynamic cases, and $\mu_s, \mu_d$ are the friction coefficients. It is straightforward to observe that, in the characterization itself, it is assumed that $\mathbf{v}_P = \pm \|\mathbf{v}_P\|\mathbf{e}_x$ is tangent to the plane, thereby excluding detachment from or impact with the plane, phenomena that must be analyzed separately.

The usual analysis of the system thus imposes the cardinal equations of Dynamics:
\begin{eqnarray*}
	\left\{
	\begin{array}{l}
		m {\mathbf{g}} + {\mathbf{\Phi}} \= m {\mathbf{a}}_G \\
		(P-G) \wedge {\mathbf{\Phi}} \= \dfrac{d}{d t} {\mathbf{\Gamma}}_G
	\end{array}
	\right.
\end{eqnarray*}
together with the constitutive relations (\ref{AC}) and the complementarity condition $y_P \Phi_y = 0$ with $y_P \ge 0, \,\Phi_y \ge 0$.

It is not the purpose of this work to analyze in detail the results of this classical approach to the study of the Painlev\'e system. For a thorough treatment, among the extensive literature, see e.g. \cite{GenotBrogliato1999, Stewart2000, zhao2015asymptotic} and the survey \cite{champneys2016painleve}. For the purposes of the present work, it suffices to recall that, by employing the standard approach, one can prove that, for suitable (though possibly large) values of the friction coefficient, there exist initial data for which the cardinal equations admit either no solution or more than one. The existence of a paradox thus becomes evident.

	\section{The geometric approach to the Painlev\'e system}
	\subsection{Geometry}

	The mechanical system underlying the Painlev\'e paradox can be modelled within the geometric framework of Regular Geometric Impulsive Mechanical Systems (RGIMS). Referring the reader to \cite{Pasquero2018RMUP,PasqueroToAppear} for a more detailed treatment of RGIMS, in the specific case of the Painlev\'e system the corresponding RGIMS is determined by the 5--tuple $(\M, \Phi, \S, \A, \B)$
	where the structures are defined as follows.

	\begin{itemize}
		\item $\M$ is the {\it configuration space--time bundle} $t : \M \to \Reali$, locally described by the coordinates $(t, x,y, \vth)$. It naturally induces the vector bundle of vertical vectors
		$V(\M) = \{ \mathbf{X} \in T(\M) | < \mathbf{X} , d t > \= 0\}$
		and the affine bundle of absolute velocities
		$\J = \{ \mathbf{X} \in T(\M) | < \mathbf{X} , d t > \= 1\}$
		modelled on $V(\M)$. Both bundles are locally described by the coordinates
		$(t, x,y, \vth,\dot{x}, \dot{y}, \dot{\vth})$.
		\item $\Phi$ is the vertical metric $\Phi:V(\M)\times V(\M)\to\Reali$ given locally by the mass matrix
		$g_{ij}=\mathrm{diag}(m,m,A)$.
		\item $\S\hookrightarrow \M$ is the subbundle of configurations satisfying the contact condition between the endpoint $P$ of the rod and the horizontal line, then called the {\it contact bundle}. For $\vth\in(0,\pi)$ its Cartesian description is $
		\S=\{y-L\sin\vth=0\}$.
		The fibration $t:\S\to\Reali$ induces the subbundles $V(\S) \subset V(\M)$ and $\JS \subset \J$.
		\item $A\hookrightarrow \J$ is the subbundle representing permanent kinematic constraints acting on the rod. In this case no such constraints are present, hence $A=\J$.
		\item $\B\hookrightarrow \JS$ is the subbundle representing instantaneous kinetic constraints modelling the roughness of the guide, then called the {\it friction bundle}. Recalling (\ref{velPuntoContatto}), it is described by the condition $\B=\{\dot x - L\dot\vth\sin\vth=0\}$.
		Since this equation is linear in the velocities, the differences of elements of $\B$ form a vector subspace	$V(\B)\subset V(\S)$.
	\end{itemize}

	From now on, we will use bold capital letters to denote elements of $V(\M), \J$ and bold lowercase letters to denote vectors in the usual euclidean space. Moreover, later on we will consider only elements ${\mathbf X} \in V(\M)$ and $\p \in \J$ that satisfy the contact condition with $\S$, i.e.\ those for which $\pi({\mathbf X}) \in \S, \, \pi({\p}) \in \S$, where $\pi:T(\M) \to \M$ is the standard projection.

	\medskip

	A generic absolute velocity $\p\in \J$ is written as
	$$
	{\p} \= \dfrac{\partial}{\partial t} \, + \, \dot{x} \, \dfrac{\partial}{\partial x} \, + \, \dot{y} \, \dfrac{\partial}{\partial y}\, + \, \dot{\vth} \, \dfrac{\partial}{\partial \vth} \, .
	$$
	A generic velocity tangent to $\S$, viewed as an element ${\p}^{\|}_{\S} \in \JS \subset \J$, has the form
	$$
	{\p}^{\|}_{\S} \= \dfrac{\partial}{\partial t} \, + \, \dot{x} \, \dfrac{\partial}{\partial x} \, + \, {L \dot{\vth} \cos \vth} \, \dfrac{\partial}{\partial y}\, + \, \dot{\vth} \, \dfrac{\partial}{\partial \vth} \, .
	$$
	A velocity tangent to $\S$ and satisfying the condition determined by $\B$,  viewed as an element ${\p}^{\|}_{\B} \in \JS \subset \J$, has the form
	$$
	{\p}^{\|}_{\B} \= \dfrac{\partial}{\partial t} \, + \, L \dot{\vth} \sin\vth \, \dfrac{\partial}{\partial x} \, + \, {L \dot{\vth} \cos \vth} \, \dfrac{\partial}{\partial y}\, + \, \dot{\vth} \, \dfrac{\partial}{\partial \vth} \, .
	$$
	The vertical metric $\Phi$ induces the decompositions
	\begin{subequations}\label{Decomp}
		\begin{eqnarray}\label{DecompV}
			V(\M) \= V(\S) \oplus V^{\perp}_{\S}(\M) \= V(\B) \oplus V^{\perp}_{\B}(\S) \oplus V^{\perp}_{\S}(\M)
		\end{eqnarray}   \vskip-0.7truecm
		\begin{eqnarray}\label{DecompJ}
			\hskip -0.6truecm \J \= \JS \oplus V^{\perp}_{\S}(\M) \= \B \oplus V^{\perp}_{\B}(\S) \oplus V^{\perp}_{\S}(\M) \, ,
		\end{eqnarray}
	\end{subequations}
	with $V(\B), V^{\perp}_{\B}(\S), V^{\perp}_{\S}(\M)$ mutually orthogonal.

	\smallskip

	In the Painlev\'e mechanical system, both vector bundles $V^{\perp}_{\S}(\M)$ and $ V^{\perp}_{\B}(\S)$ have a $1$--dimensional fiber. It follows from a straightforward calculation that
	$V^{\perp}_{\S}(\M) \= Lin\{{\mathbf K}_{\S}\}, \,  V^{\perp}_{\B}(\S) \= Lin\{{\mathbf K}_{\B}\} $,
	where ${\mathbf K}_{\S}, {\mathbf K}_{\B}$ are the unit vectors
	\begin{subequations}\label{GeneratoriOrtogonali}
		\begin{eqnarray}\label{GeneratoreVS}
			\hskip-0.6truecm		\begin{array}{lcl}
				{\mathbf K}_{\S} &=& \sqrt{\dfrac{mL^2 \cos^2 \vth +A}{mA}} \, \left(\dfrac{A}{mL^2 \cos^2 \vth +A} \, \dfrac{\partial}{\partial y} \right. \\ \\
				&& \hskip4truecm \left. - \, \dfrac{mL \cos \vth}{mL^2 \cos^2 \vth +A}  \,  \dfrac{\partial}{\partial \vth} \right) \, ;
			\end{array}
		\end{eqnarray}
		\begin{eqnarray}\label{GeneratoreVB}
			\begin{array}{lcl}
				{\mathbf K}_{\B} &=& \sqrt{\dfrac{mL^2 + A}{m(mL^2 \cos^2 \vth +A)}} \, \left(\dfrac{mL^2 \cos^2\vth +A}{mL^2 +A} \, \dfrac{\partial}{\partial x} \right. \\ \\
				&& \left. \hskip 2truecm
				- \, \dfrac{mL^2 \sin\vth \cos\vth}{mL^2 +A}  \, \dfrac{\partial}{\partial y}
				- \, \dfrac{mL \sin\vth}{mL^2 +A} \,  \dfrac{\partial}{\partial \vth} \right) \, .
			\end{array}
		\end{eqnarray}
	\end{subequations}

	In particular, using (\ref{Decomp},\ref{GeneratoriOrtogonali}), a generic velocity $\p$ may then be written in the following forms
	\begin{itemize}
		\item if we consider $\p \= \p^{\|}_{\S}(\p) \, + \, {\mathbf V}^{\perp}_{\S}(\p)$, then
		\begin{subequations}
			\begin{eqnarray}\label{DecompVp1}
				\begin{array}{lcl}
					\p^{\|}_{\S}(\p) =  \dfrac{\partial}{\partial t}  &+&  \dot{x} \, \dfrac{\partial}{\partial x} \\ \\ &+& \left(\dfrac{mL^2\cos^2\vth}{mL^2\cos^2\vth +A} \, \dot{y} \, + \,  \dfrac{AL\cos\vth}{mL^2\cos^2\vth +A} \, \dot{\vth}\right)  \, \dfrac{\partial}{\partial y} \\ \\ &+& \left(\dfrac{mL\cos\vth}{mL^2\cos^2\vth +A} \, \dot{y} \, + \,  \dfrac{A}{mL^2\cos^2\vth +A} \, \dot{\vth}\right) \, \dfrac{\partial}{\partial \vth}
				\end{array}
			\end{eqnarray}
			\begin{eqnarray}\label{DecompJp1}
				\hskip-1.2truecm	\begin{array}{lcl}
					{\mathbf V}^{\perp}_{\S}(\p) &\=& \sqrt{\dfrac{mA}{mL^2 \cos^2 \vth +A}} \left(\dot{y} - L \dot{\vth} \cos \vth\right) {\mathbf K}_{\S} \\ \\
					&\= & \dfrac{A}{mL^2 \cos^2 \vth +A} \,\left(\dot{y} - L \dot{\vth} \cos \vth\right) \, \dfrac{\partial}{\partial y} \\ \\
					&&  - \, \dfrac{mL \cos \vth}{mL^2 \cos^2 \vth +A} \,\left(\dot{y} - L \dot{\vth} \cos \vth\right) \,  \dfrac{\partial}{\partial \vth}
				\end{array}
			\end{eqnarray}
		\end{subequations}
		\item if we consider $\p \= {\mathbf p}^{\|}_{\B}(\p) \, + \, {\mathbf V}^{\perp}_{\B}(\p) \, + \, {\mathbf V}^{\perp}_{\S}(\p)$, then
		\begin{subequations}
			\begin{eqnarray}\label{DecompVp2}
				\hskip-2.4truecm	\begin{array}{lcl}
					\p^{\|}_{\B}(\p) =  \dfrac{\partial}{\partial t}  &+&  \dfrac{L \sin \vth}{mL^2 +A} \left(mL\dot{x} \sin \vth + mL\dot{y} \cos \vth + A \dot{\vth}\right) \, \dfrac{\partial}{\partial x} \\ \\  &+& \dfrac{L \cos \vth}{mL^2 +A} \left(mL\dot{x} \sin \vth + mL\dot{y} \cos \vth + A \dot{\vth}\right) \, \dfrac{\partial}{\partial y} \\ \\ &+& \dfrac{1}{mL^2 +A} \left(mL\dot{x} \sin \vth + mL\dot{y} \cos \vth + A \dot{\vth}\right) \, \dfrac{\partial}{\partial \vth}
				\end{array}
			\end{eqnarray}
			\begin{eqnarray}\label{DecompJpB2}
				\hskip-1.2truecm \begin{array}{lcl}
					{\bf V}^{\perp}_{\B}(\p) &=&  \sqrt{\dfrac{m(mL^2 \cos^2 \vth +A)}{mL^2 + A}} \left[\dot{x} - \dfrac{L \sin \vth}{mL^2 \cos^2 \vth +A} \left(mL\dot{y} \cos \vth + A \dot{\vth}\right)\right]   {\mathbf K}_{\B} \\ \\
					&\= & \dfrac{mL^2 \cos^2\vth +A}{mL^2 +A} \left[\dot{x} - \dfrac{L \sin \vth}{mL^2 \cos^2 \vth +A} \left(mL\dot{y} \cos \vth + A \dot{\vth}\right)\right] \, \dfrac{\partial}{\partial x} \\ \\
					&& - \, \dfrac{mL^2 \sin\vth \cos\vth}{mL^2 +A} \left[\dot{x} - \dfrac{L \sin \vth}{mL^2 \cos^2 \vth +A} \left(mL\dot{y} \cos \vth + A \dot{\vth}\right)\right] \, \dfrac{\partial}{\partial y} \\ \\
					&& - \, \dfrac{mL \sin\vth}{mL^2 +A} \left[\dot{x} - \dfrac{L \sin \vth}{mL^2 \cos^2 \vth +A} \left(mL\dot{y} \cos \vth + A \dot{\vth}\right)\right] \,  \dfrac{\partial}{\partial \vth}
				\end{array}
			\end{eqnarray}
			\begin{eqnarray}\label{DecompJp2}
				\hskip-2.4truecm	\begin{array}{lcl}
					{\mathbf V}^{\perp}_{\S}(\p) &\=& \sqrt{\dfrac{mA}{mL^2 \cos^2 \vth +A}} \left(\dot{y} - L \dot{\vth} \cos \vth\right) {\mathbf K}_{\S} \\ \\
					&\= & \dfrac{A}{mL^2 \cos^2 \vth +A} \,\left(\dot{y} - L \dot{\vth} \cos \vth\right) \, \dfrac{\partial}{\partial y} \\ \\
					&&  - \, \dfrac{mL \cos \vth}{mL^2 \cos^2 \vth +A} \,\left(\dot{y} - L \dot{\vth} \cos \vth\right) \,  \dfrac{\partial}{\partial \vth}
				\end{array}
			\end{eqnarray}
		\end{subequations}
	\end{itemize}
	The above-mentioned decompositions, through the use of the vectors ${\mathbf K}_{\S}, {\mathbf K}_{\B}$, allow one to formalize the unilateral nature of the constraint system and to determine whether and when a generic velocity $\p$ gives rise to an impact:
	\begin{itemize}
		\item the condition ${\mathbf \Phi}({\mathbf V}^{\perp}_{\S}(\p), {\mathbf K}_{\S}) < 0$  implies that the rod impacts the rough line. Conversely, the condition ${\mathbf \Phi}({\mathbf V}^{\perp}_{\S}(\p), {\mathbf K}_{\S}) > 0$  is indicative of a detachment of the point $P$ of the rod from the rough line;
		\item it is also possible to require the unilaterality of the constraint $\B$ for instance expressed by the condition  ${\mathbf \Phi}({\mathbf V}^{\perp}_{\B}(\p), {\mathbf K}_{\B}) > 0$.  This form of unilaterality becomes particularly significant when ${\mathbf \Phi}({\mathbf V}^{\perp}_{\S}(\p), {\mathbf K}_{\S}) = 0$.
	\end{itemize}
	These conditions will be encountered again in the following subsection, which is devoted to the constitutive characterization of the constraint.

	\medskip

	Taking into account (\ref{velPuntoContatto}) and with obvious notation, the initial condition requiring the tangency of the contact point $P$ to the constraint at $t_0$ is given by the condition $\dot y_0 - L\dot\vth_0\cos\vth_0 = 0$. Then, if $\pl = \p(t_0)$, the initial data of the Painlev\'e paradox are such that
	\begin{itemize}
		\item if we split $\pl$ as $\pl \= \p^{\|}_{\S}(\pl) \, + \, {\mathbf V}^{\perp}_{\S}(\pl)$, then
		\begin{eqnarray}\label{DecompVp1-1}
			\begin{array}{l}
				\p^{\|}_{\S}(\pl) =  \dfrac{\partial}{\partial t}  +  \dot{x}_0 \, \dfrac{\partial}{\partial x} + L\dot{\vth}_0 \cos\vth_0  \, \dfrac{\partial}{\partial y} + \dot{\vth}_0\, \dfrac{\partial}{\partial \vth} \\ \\
				{\mathbf V}^{\perp}_{\S}(\pl) \= 0
			\end{array}
		\end{eqnarray}
		\item if we split $\pl$ as $\pl \= \p^{\|}_{\B}(\pl) \, + \, {\mathbf V}^{\perp}_{\B}(\pl) \, + \, {\mathbf V}^{\perp}_{\S}(\pl)$, then
		\begin{eqnarray}\label{DecompVp2-2}
			\hskip-2truecm	\begin{array}{lcl}
				\p^{\|}_{\B}(\pl) &=&  \dfrac{\partial}{\partial t}  + L\sin\vth_0 \left[\dot{\vth}_0 + \dfrac{mL \sin \vth_0}{mL^2 +A} \left(\dot{x}_0 - L \dot{\vth}_0 \sin\vth_0 \right) \right] \, \dfrac{\partial}{\partial x} \\ \\ && \qquad + L\cos\vth_0 \left[\dot{\vth}_0 + \dfrac{mL \sin \vth_0}{mL^2 +A} \left(\dot{x}_0 - L \dot{\vth}_0 \sin\vth_0 \right) \right] \, \dfrac{\partial}{\partial y} \\ \\ && \qquad \quad + \left[\dot{\vth}_0 + \dfrac{mL \sin \vth_0}{mL^2 +A} \left(\dot{x}_0 - L \dot{\vth}_0 \sin\vth_0 \right) \right] \, \dfrac{\partial}{\partial \vth}
				\\ \\
				{\bf V}^{\perp}_{\B}(\pl) &=&  \dfrac{mL^2 \cos^2\vth_0 +A}{mL^2 +A} \left(\dot{x}_0 - L \dot{\vth}_0 \sin\vth_0 \right) \, \dfrac{\partial}{\partial x}  \\ \\
				&& - \, \dfrac{mL^2 \sin\vth_0 \cos\vth_0}{mL^2 +A} \left(\dot{x}_0 - L \dot{\vth}_0 \sin\vth_0 \right) \, \dfrac{\partial}{\partial y} \\ \\
				&& - \, \dfrac{mL \sin\vth_0}{mL^2 +A} \left(\dot{x}_0 - L \dot{\vth}_0 \sin\vth_0 \right) \,  \dfrac{\partial}{\partial \vth}
				\\ \\
				{\mathbf V}^{\perp}_{\S}(\pl) &=& 0
			\end{array}
		\end{eqnarray}
	\end{itemize}

	The decompositions (\ref{DecompVp1-1}, \ref{DecompVp2-2}) reveal the first novel result of this work. Their analysis makes it immediately apparent that, in the classical modelling of the Painlev\'e system, where friction is treated as a force and therefore is not regarded as a geometric component of the framework, the initial data of the problem determine a system velocity that is tangent to the only constraint present, namely the positional constraint $\S$ defined by the rough line. As a consequence, the kinematics of the system does not exhibit any impulsive behaviour, which is instead clearly observed in experimental data (see, for instance, \cite{Zhao2008Experimental}). It thus becomes necessary to introduce the notion of ``impact without collision'', which has been described in various ways in the literature (see e.g. \cite{GenotBrogliato1999, champneys2016painleve}).

	By contrast, in the geometric modelling in which friction is treated as an instantaneous kinematic constraint $\B$ internal to the positional constraint $\S$, the decomposition shows, consistently with the previous case, a vanishing component orthogonal to $\S$, but also a non-vanishing component orthogonal to $\B$, whose magnitude depends on the orientation angle of the rod and on the sliding velocity $\dot{x}_0 - L \dot{\vth}_0 \sin\vth_0$ of the contact point $P$ along the line. This implies that, if such a sliding velocity is non-zero, an impact occurs within $\S$ against the kinematic constraint $\B$ arising from the modelling of friction. The system therefore exhibits an impulsive character, thereby determining the (impulsive) nature of the constitutive characterization of the constraint, which is the subject of the following subsection.

	\subsection{Constitutive characterization}

	Once the Painlev\'e paradox has been framed within the theory of systems subject
	to impulsive constraints, the restoration of determinism can be achieved by
	assigning an impulsive-type constitutive characterization, namely (see~\cite{Pasquero2005uni, Pasquero2008, Pasquero2018RMUP}) by
	introducing a rule ${\mathbf I}: \J \to V(\M)$ such that, for every incoming velocity $\pl\in\J$ that gives rise to an
	impact, one can uniquely assign, once the geometric and kinematic data at the
	impact time $t_0$ are known, an outgoing velocity
	$\pr  = \pl + \mathbf{I}(\pl) \in \J$
	of the system, which obviously does not in turn generate another impact.
	Moreover, although any rule with this property is mathematically admissible in
	that it restores determinism, a constitutive characterization should also
	provide outgoing velocities that are as consistent as possible with
	experimental observations.

	Referring once again the reader to \cite{Pasquero2018RMUP} for a more detailed discussion of the concept
	of constitutive characterization in the framework of RGIMS, the most general
	constitutive characterization for the Painlev\'e system is of the form
	\begin{eqnarray}\label{CarCostGenerale}
		\mathbf{I}(\pl) \= \sigma \, \mathbf{K}_{\S} \, + \, \beta \, \mathbf{K}_{\B}
	\end{eqnarray}
	where $\sigma$ and $\beta$ are coefficients depending on the mass, geometric,
	and kinematic data of the system at the impact instant.

	With regard to the contact constraint $\S$, unilaterality must be imposed, and
	the following conditions are adopted:
	\begin{itemize}
		\item the condition ${\mathbf \Phi}({\mathbf V}^{\perp}_{\S}(\p), {\mathbf K}_{\S}) < 0$ is always regarded as an impact
		condition with the unilateral contact constraint $\S$;
		\item the condition ${\mathbf \Phi}({\mathbf V}^{\perp}_{\S}(\p), {\mathbf K}_{\S}) > 0$ is always regarded as a detachment condition from $\S$, and therefore determines an outgoing velocity;
		\item the condition ${\mathbf \Phi}({\mathbf V}^{\perp}_{\S}(\p), {\mathbf K}_{\S}) = 0$ may determine either an impact or an outgoing velocity only by considering the additional condition
		${\mathbf \Phi}({\mathbf V}^{\perp}_{\B}(\p), {\mathbf K}_{\B}) \lesseqgtr 0$.
	\end{itemize}

	Recalling that, since ${\mathbf V}^{\perp}_{\S}(\pl) =0$, in the Painlev\'e problem one always has
	${\mathbf \Phi}({\mathbf V}^{\perp}_{\S}(\pl), {\mathbf K}_{\S}) = 0$, it follows that:
	\begin{itemize}
		\item the condition $\sigma>0$ always determines an outgoing velocity;
		\item the condition $\sigma<0$ never determines an outgoing velocity.
	\end{itemize}

	With regard to the friction constraint $\B$, unilaterality is not mandatory, and
	the following cases arise:
	\begin{itemize}
		\item the unilaterality of $\B$ can be expressed in the form
		${\mathbf \Phi}({\mathbf V}^{\perp}_{\B}(\p), {\mathbf K}_{\B}) > 0$, which determines an impact whenever
		${\mathbf u}\cdot {\mathbf v}_P <0$ (see Fig.~1), that is, when the rod is pushed
		to move while remaining in contact with the rough line. In this case, the
		condition ${\mathbf \Phi}({\mathbf V}^{\perp}_{\B}(\p), {\mathbf K}_{\B}) \le 0$ does not determine an impact, since the
		contact point $P$ is either at rest relative to the rough line or the rod is
		pulled while moving in contact with it;
		\item alternatively, the constraint $\B$ may be considered bilateral, so that an
		impact occurs whenever ${\mathbf \Phi}({\mathbf V}^{\perp}_{\B}(\p), {\mathbf K}_{\B}) \ne 0$.
	\end{itemize}

	Once these preliminary conditions have been established, the determination of
	a constitutive characterization, namely of the coefficients $\sigma$ and
	$\beta$, may be based on criteria of simplicity and analogy with other impact
	models. Starting from the incoming velocity
	$\pl \= \dfrac{\partial}{\partial t}  +  \dot{x}_0 \, \dfrac{\partial}{\partial x} + L\dot{\vth}_0 \cos\vth_0  \, \dfrac{\partial}{\partial y} + \dot{\vth}_0\, \dfrac{\partial}{\partial \vth} \in \JS$
	we list below several possible choices.

	\subsubsection{Example 1: rebound or stop}

	Assume both $\S$ and $\B$ to be unilateral. The system impacts the friction
	constraint if and only if ${\mathbf \Phi}({\mathbf V}^{\perp}_{\B}(\pl), {\mathbf K}_{\B}) > 0$. By assigning the outgoing
	conditions
	\begin{eqnarray}\label{CondUScita}
		{\mathbf \Phi}({\mathbf V}^{\perp}_{\S}(\pr), {\mathbf K}_{\S}) > 0 \quad \textrm{or} \quad
		\left\{ \begin{array}{l}
			{\mathbf \Phi}({\mathbf V}^{\perp}_{\S}(\pr), {\mathbf K}_{\S}) = 0 \\
			{\mathbf \Phi}({\mathbf V}^{\perp}_{\B}(\pr), {\mathbf K}_{\B}) \le 0
		\end{array} \right.
	\end{eqnarray}
	one may set $\sigma = 0, \, \beta = -(1+\varepsilon)\| {\bf V}^{\perp}_{\B}(\pl) \|, \, \varepsilon \in (0,1)$, obtaining
	$\pr \= \p^{\|}_{\B}(\pl) \, - \, \varepsilon {\bf V}^{\perp}_{\B}(\pl) $.
	In this case, the rod does not impact $\B$ if it is pulled, and
	rebounds against $\B$ if it is pushed against it, without detachment
	of the contact point $P$ from $\S$.

	Alternatively, by setting  $\sigma = 0, \, \beta = -\| {\bf V}^{\perp}_{\B}(\pl) \| $, one obtains
	$\pr \= \p^{\|}_{\B}(\pl)$. In this case, the rod does not impact $\B$
	if it is pulled, while if it is pushed against it, the contact point comes to
	rest and the rod begins to rotate about the contact point.

	In both cases, one may require $\sigma$ to be strictly positive, which
	implies that $\pr$ is always an outgoing velocity due to the condition
	${\mathbf \Phi}({\mathbf V}^{\perp}_{\S}(\pr), {\mathbf K}_{\S}) > 0$, and that the contact point always detaches from $\S$.

	\subsubsection{Example 2: maximum possible braking}

	Assume again both $\S$ and $\B$ to be unilateral, with outgoing conditions still
	given by (\ref{CondUScita}). Similarly to what is proposed in \cite{PasqueroToAppear}, one may set
	\begin{eqnarray}\label{FaIlPoss}
		\beta \= \left\{
		\begin{array}{lcl}
			- \, \| {\bf V}^{\perp}_{\B}(\pl) \| \quad &{\textrm{if}}& \quad \| {\bf V}^{\perp}_{\B}(\pl) \| \le \mu \\ \\
			- \, \mu \quad &{\textrm{if}}& \quad \| {\bf V}^{\perp}_{\B}(\pl) \| > \mu
		\end{array}	 \right.
	\end{eqnarray}
	where $\mu>0$ is a prescribed coefficient. Note that, in general,
	$ \| {\bf V}^{\perp}_{\B}(\pl) \|$ depends on the inclination angle $\vth_0$ of the rod at
	impact. In this case, the rod does not impact the constraint if it is pulled,
	and is braked by the constraint if it is pushed against it. If the impact
	velocity of the contact point is less than or equal to $\mu$, the contact point
	stops and the rod continues its motion by rotating about the contact point. If,
	instead, the velocity exceeds $\mu$, the constraint brakes the contact point as
	much as possible without stopping it. In this case, it is therefore necessary,
	in order to obtain an outgoing velocity, that $\sigma>0$, which entails the
	detachment of the contact point from the constraint $\S$.

	\begin{Remark}
		The possible dependence of $\sigma$ on the mass properties of the body $m,A$
		models the action of gravity and of the rod's rotational inertia in the impact
		against the friction constraint.
	\end{Remark}

	\subsubsection{Example 3: detachment in any case}

	Assume $\S$ to be unilateral and $\B$ bilateral. The impact condition becomes
	${\mathbf \Phi}({\mathbf V}^{\perp}_{\B}(\pl), {\mathbf K}_{\B}) \ne 0$, while the outgoing condition reduces to
	${\mathbf \Phi}({\mathbf V}^{\perp}_{\S}(\pr), {\mathbf K}_{\S}) > 0$, which implies $\sigma>0$. In this case, it is
	possible to differentiate the system behavior (for instance, the value of
	$\sigma$) depending on whether the rod is pushed or pulled against the friction
	constraint. One may set, for example,
	\begin{eqnarray}\label{FaIlPoss2}
		\begin{array}{l}
			\sigma \= \left\{
			\begin{array}{lcl}
				(\lambda_1 m + \alpha_1 A) \| {\bf V}^{\perp}_{\B}(\pl) \| \quad &{\textrm{if}}& \quad {\mathbf \Phi}({\mathbf V}^{\perp}_{\B}(\pl), {\mathbf K}_{\B}) > 0 \\ \\
				(\lambda_2 m + \alpha_2 A) \| {\bf V}^{\perp}_{\B}(\pl) \| \quad &{\textrm{if}}& \quad {\mathbf \Phi}({\mathbf V}^{\perp}_{\B}(\pl), {\mathbf K}_{\B}) < 0
			\end{array}	 \right. \\ \\
			\beta \= \left\{
			\begin{array}{lcl}
				- \, \| {\bf V}^{\perp}_{\B}(\pl) \| \quad &{\textrm{if}}& \quad  {\mathbf \Phi}({\mathbf V}^{\perp}_{\B}(\pl), {\mathbf K}_{\B}) > 0,  \| {\bf V}^{\perp}_{\B}(\pl) \| \le \mu \\ \\
				- \, \mu \quad &{\textrm{if}}& \quad  {\mathbf \Phi}({\mathbf V}^{\perp}_{\B}(\pl), {\mathbf K}_{\B}) > 0 , \| {\bf V}^{\perp}_{\B}(\pl) \| > \mu \\ \\
				- \, \gamma \, \| {\bf V}^{\perp}_{\B}(\pl) \| \quad &{\textrm{if}}& \quad  {\mathbf \Phi}({\mathbf V}^{\perp}_{\B}(\pl), {\mathbf K}_{\B}) < 0
			\end{array}	 \right.
		\end{array}
	\end{eqnarray}
	where $\lambda_1, \lambda_2, \alpha_1, \alpha_2, \mu, \gamma$ are positive coefficients, $\gamma \in (0,1)$.
	Such a choice on $\sigma$ differentiates the detachment of the contact point from $\S$ depending on whether the rod is pushed or pulled, and on $\beta$ combines the characterization of the maximum admissible braking  when the rod is pushed with a simple braking when the rod is pulled.

	\medskip

	In the examples described above, in analogy with the laws of dry friction, we have always considered a dependence of the coefficients on the norm of orthogonal velocities at most linear, but it is obviously possible to consider different choices, for example quadratic, especially when these norms exceed a threshold value.

	\section{Conclusions and developments}

The Painlev\'e mechanical system, together with its associated paradox, can be framed within the context of Classical Mechanics of systems with a finite number of degrees of freedom subject to unilateral constraints with friction, where it represents a challenging anomaly. Interest in this topic is both of a purely theoretical nature, attested by the extensive bibliography available on the subject, which analyzes the genuine paradox concerning the determinism of Classical Mechanics arising from the study of this system, and of an applied nature, as evidenced by the clear experimental observation of the phenomenon (see, for example, \cite{Zhao2008Experimental}), which can be detected both in laboratory settings and in engineering applications.

The alternative approach to the problem proposed here, which instead situates the Painlev\'e system within the framework of Impulsive Mechanics of systems subject to positional and kinematic constraints, exhibits several satisfactory features. First of all, by virtue of the very nature of the approach, no issues arise concerning the determinism of the system’s evolution, which is uniquely determined once the geometric and kinematic data are specified. Moreover, similarly to what occurs in the elementary model of dry friction, as well as in more advanced models of non-ideal impulsive systems (see, for instance, \cite{PasqueroToAppear}), the analysis of the evolution is reduced to the study of two numerical coefficients, denoted by $\sigma$ and $\beta$, which depend on the geometric, kinematic, and possibly dynamic data of the problem (as well as on temperature, materials, and other factors), and which enter into the description of the impact. Finally, this framework allows for an experimental analysis of the system’s behavior using relatively simple instrumentation. For example, with reference to the methodologies presented in \cite{Zhao2008Experimental}, replacing the moving belt or moving rail with an analogous belt or rail that is smooth over one portion and rough over another would make it possible to experimentally determine the coefficients $\sigma$ and $\beta$, with their dependence on geometric and kinematic data (as well as dynamic parameters such as the weight and moment of inertia of the rod, and other factors not necessarily of a mechanical nature, such as temperature and materials).

\medskip

The possible developments of such an approach are manifold. First of all, the generality of the method allows its application to the analysis of mechanical systems other than the Painlev\'e one; for instance, one may consider a mobile disk in contact with a horizontal plane that is partially smooth over one region and partially rough over another.

Moreover, the underlying idea of the method does not require the system to be planar. By the same methodology, it is possible to investigate the three-dimensional Painlev\'e problem, in which the rod moves in three-dimensional space with one of its ends in contact with a rough horizontal plane.

Finally, the approach provides a natural framework for the study of so-called inert constraints (see, for example, the classical book \cite{Pars}) and, consequently, of variable-topology mechanical systems, together with their applications in robotics and biomechanics (see, for instance, \cite{Kovecses2009,Carpentier2010} and the references therein).

	\bibliographystyle{unsrt}
	\bibliography{bibart,biblib}

\end{document}